\documentclass[aps,prb,reprint,superscriptaddress,showpacs]{revtex4-1}
\usepackage{graphicx,amsmath,amssymb,latexsym}
\usepackage{txfonts}
\DeclareSymbolFont{symbols}{OMS}{cmsy}{m}{n}
\DeclareSymbolFont{largesymbols}{OMX}{cmex}{m}{n}
\renewcommand{\bm}[1]{\boldsymbol #1}

\begin{document}

\title{
Repulsion-to-attraction transition in correlated electron systems triggered by a monocycle pulse
}

\author{Naoto Tsuji}
\affiliation{Institut f\"{u}r Theoretische Physik, ETH Zurich, 8093 Z\"{u}rich, Switzerland}
\author{Takashi Oka}
\affiliation{Department of Physics, University of Tokyo, Hongo, Tokyo 113-0033, Japan}
\author{Hideo Aoki}
\affiliation{Department of Physics, University of Tokyo, Hongo, Tokyo 113-0033, Japan}
\author{Philipp Werner}
\affiliation{Institut f\"{u}r Theoretische Physik, ETH Zurich, 8093 Z\"{u}rich, Switzerland}
%\email[]{}
%\homepage[]{}
%\thanks{}
%\altaffiliation{}

\begin{abstract}
We study the time evolution of the Hubbard model driven by a half-cycle or monocycle pulsed electric field $F(t)$
using the nonequilibrium dynamical mean-field theory. We find that for properly chosen pulse shapes
the electron-electron interaction can be effectively and permanently switched from repulsive to attractive
if there is no energy dissipation.
The physics behind the interaction conversion is a nonadiabatic shift $\delta$ of the population in momentum space. 
When $\delta\sim\pi$, the shifted population relaxes to a negative-temperature state, which leads to the interaction switching.
Due to electron correlation effects  
$\delta$ deviates from the dynamical phase $\phi=\int dt F(t)$, which enables 
the seemingly counterintuitive 
repulsion-to-attraction transition by a monocycle pulse with $\phi=0$.
\end{abstract}

%\collaboration{}
%\noaffiliation

%03.75.Lm    : Tunneling, Josephson effect, Bose-Einstein condensates in periodic potentials, 
%              solitons, vortices, and topological excitations
%03.75.Ss    : Matter waves
%                --- Degenerate Fermi gases 
%03.75.-b    : Matter waves
%7.85.-d     : Ultracold gases, trapped gases
%67.85.Lm    : Degenerate Fermi gases 
%03.65.Vf    : Quantum mechanics
%                --- Phases: geometric; dynamic or topological 
%71.10.Fd    : Theories and models of many-electron systems
%                --- Lattice fermion models (Hubbard model, etc.)
%05.70.Ln    : Thermodynamics
%                --- Nonequilibrium and irreversible thermodynamics
%05.10.Ln    : Computational methods in statistical physics and nonlinear dynamics
%                --- Monte Carlo methods
%71.27.+a   : Strongly correlated electron systems; heavy fermions
%05.30.Fk   : Quantum statistical mechanics 
%               --- Fermion systems and electron gas

\date{\today}

\maketitle

\section{Introduction}

Controlling the interparticle interactions in a correlated electron system by 
applying intense laser fields 
is a challenging and exciting perspective, which may lead to states of matter that do not exist in equilibrium.   
For example, if one could effectively change the electron-electron interaction from the original Coulomb repulsion to an attraction, 
this may induce an $s$-wave superconducting state with very high transition temperature (optimally $\approx 0.1$ bandwidth),
\cite{MicnasRanningerRobaszkiewicz1990,KellerMetznerSchollwoeck2001}
or the BCS-BEC crossover. \cite{NozieresSchmittRink1985,KellerMetznerSchollwoeck2001}
It will also enable us to study phenomena characteristic of 
nonequilibrium quantum systems, such as transient states after an interaction quench.
\cite{ManmanaWesselNoackMuramatsu2007,MoeckelKehrein2008,EcksteinKollarWerner2009}
The control of interparticle interactions is in fact possible in
cold-atom systems, \cite{BlochDalibardZwerger2008}
where one can manipulate the interaction in a wide range from repulsive to attractive using 
the Feshbach resonance that dominates the scattering 
length, \cite{ChinGrimmJulienneTiesinga2010} but 
such a technique cannot be applied to electron systems. 

One way to control the interaction is to create a population inversion in metallic bands corresponding to
a negative-temperature ($T$) state. \cite{PurcellPound1951,Ramsey1956}
This implies an effective switching of the interaction from repulsive to attractive,
since a density matrix $e^{-H/T}$ for a Hamiltonian $H$ with temperature $T<0$ corresponds to the one
for the inverted $-H$ with $-T>0$. \cite{RappMandtRosch2010,TsujiOkaWernerAoki2011}
While a (partial) population inversion itself is a common phenomenon (e.g., in laser productions),
the interaction conversion is a genuine correlation effect in nonequilibrium. 
Ideally, the laser fields that drive the system should be ``pulsed'' waves
since, first, the available intensity is 
generally much higher for ultrafast pulses
\cite{Cavalieri2007,Wall2011,UlbrichtHendryShanHeinz2011} than for continuous-wave lasers, 
and second, continued heating can be avoided. These considerations raise a fundamental question: 
can irradiation by a single-cycle pulse put a system into a negative-$T$ state that survives for a long time after the pulse?

In this paper, we show that it is possible to induce a population inversion in metallic systems using a properly shaped monocycle or half-cycle pulse, and that in the absence of energy dissipation, the system will thermalize in the negative-$T$ state after the pulse. 
By solving the driven Hubbard model 
with the nonequilibrium dynamical mean-field theory (DMFT), \cite{GeorgesKotliarKrauthRozenberg1996,FreericksTurkowskiZlatic2006}
we will demonstrate that pulse fields $F(t)$ with proper asymmetry
between the positive [$F(t)>0$] and negative [$F(t)<0$] parts
trigger a repulsion-to-attraction transition.
Such asymmetric pulses can readily be generated thanks to the recent progress in laser techniques, 
\cite{JonesYouBucksbaum1993,Hebling2008,JawariyaNagaiTanaka2009}
while their potential application to correlated systems has remained unexplored, 
in contrast to symmetric, many-cycle pulses. \cite{MoritzDevereauxFreericks2010,EcksteinWerner2011a}

%If we depart from the Floquet states to consider monocycle pulses, we can immediately raise  two questions.  
%First, can a single pulse (with an electric field with $\int F(t)=0$ 
%as dictated by Maxwell's equation) trigger such an effect? Second, will the effect of the pulse persist? 
%Here we show that due to correlation effects, 
%a monopulse, when asymmetric with different shapes for the positive ($F(t)>0$) and negative ($F(t)<0$) parts, 
%a monocycle pulse can indeed produce a repulsion-to-attraction conversion, and that in an isolated system, the effect does persist after the pulse.
%Monocycle pulses with the desired properties (asymmetric shapes) can readily be generated 
%thanks to the recent progress in laser techniques \cite{JonesYouBucksbaum1993,Hebling2008,JawariyaNagaiTanaka2009}, 
%while their potential application to correlated systems has remained unexplored, 
%in contrast to symmetric, many-cycle pulses \cite{MoritzDevereauxFreericks2010,EcksteinWerner2011a}.

\begin{figure}[b]
\includegraphics[width=8.5cm]{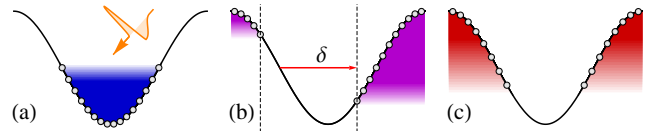}
\caption{(Color online) Schematic band pictures for the pulse-induced phase shift. 
(a) Initial system in equilibrium, 
(b) right after the pulse excitation, with the population shifted in momentum space by $\delta$, and 
(c) the system finally thermalized with a negative temperature when $\delta\simeq\pi$.}
\label{band shifting}
\end{figure}
Our strategy is to induce a nonadiabatic shift (denoted by $\delta$) 
in the momentum distribution of the electrons by the asymmetric monocycle pulse (see Fig.~\ref{band shifting}).
If we can achieve $\delta\simeq\pi$ (half of the Brillouin zone), 
the system is brought to a negative-$T$ state [Fig.~\ref{band shifting}(c)], which
amounts to a change of the interaction from repulsive to attractive. %as discussed in Ref.~\onlinecite{TsujiOkaWernerAoki2011}.
In a one-body picture, one expects that each electron acquires from the pulse field a dynamical phase
\begin{align}
\phi = \frac{ea}{\hslash}\int_{-\infty}^{\infty} dt F(t)
\label{dynamical phase}
\end{align}
with $e$ the elementary charge and $a$ the lattice constant
(hereafter we set $e=a=\hslash=1$).
This causes a momentum shift $k\to k+\phi$, so that we simply have $\delta=\phi$.
An immediate question is: can a monocycle pulse with $\int dt F(t)=0$ 
(as dictated by Maxwell's equation \cite{half-cycle}) 
%still trigger such an effect? 
induce a nontrivial shift of the population?
%Second, will the effect of the pulse persist?
We show that it is in fact possible in interacting systems,
%For interacting systems, however, this is no longer the case: 
%as we shall explicitly demonstrate, the nonadiabatic shift $\delta$ for the driven Hubbard model, solved with the nonequilibrium dynamical mean-field theory (DMFT)
%\cite{GeorgesKotliarKrauthRozenberg1996,FreericksTurkowskiZlatic2006}, 
where the nonadiabatic shift $\delta$ exhibits a clear deviation from $\phi$ due to correlation effects.
This allows us to achieve $\delta\simeq\pi$ even when $\int dt F(t)\propto\phi=0$ by choosing the pulse shape appropriately.
%We show that even when $\int dt F(t) \propto \phi=0$ (as is usually the case for pulsed laser fields \cite{half-cycle}) 
%we can achieve $\delta\simeq\pi$ 
%by choosing the pulse shape appropriately.  
%In this case the system is brought to a negative-$T$ state [Fig.~\ref{band shifting}(c)], which
%amounts to a change of the interaction from repulsive to attractive as discussed in Ref.~\onlinecite{TsujiOkaWernerAoki2011}, and numerically 
%demonstrated through the time-evolution of the double occupancy.
We reveal conditions for the pulse shape that lead to the population inversion,
and construct a `phase diagram' for the pulse-driven Hubbard model.
%A totally different mechanism from the Floquet-state situation 
%most clearly appears in the fact 
We emphasize that the interaction conversion robustly 
persists after the pulse has passed, at least in an isolated system without energy dissipation.  
This contrasts with the previously proposed scenario for the repulsion-to-attraction transition using continuous-wave fields. \cite{TsujiOkaWernerAoki2011}
%Experimentally, the pulse-induced repulsion-to-attraction conversion should be easier to detect with time-resolved spectroscopy.
%We discuss an experimental feasibility of the repulsion-to-attraction transition proposed here, and propose some
%materials which may be suitable for it.

%A common intuition is that feature of pulse fields that are normally employed in experiments is that 
%the integral of the pulse shape over a whole time domain vanishes, i.e., a quantity
%with $F(t)$ the pulse field becomes zero.
%On the other hand, there is another class of pulses, a so-called half-cycle pulse \cite{JonesYouBucksbaum1993,Hebling2008}
%that has only a half cycle of oscillations in the pulse duration. What is characteristic to it is 
%that it has nonzero $A$.

%An ultrafast pulse excitation of correlated electron systems is one of main interests in
%recent experiments of a time-resolved optical spectroscopy measurement \cite{Iwai2003a,Okamoto2007a,Wall2011}. 
%It has proven to be quite useful to gain a physical insight that has been hard to obtain in equilibrium, 
%(e.g., carrier relaxations, electron-phonon coupling, energy dissipation, etc.), 
%and also to be even possible to generate physical properties that have not appeared in equilibrium \cite{Ichikawa2011}.
%their electronic structures dynamically and drastically.
%Here we would like to pose a challenging issue: is there any way to control the many-body interaction of electrons?
%If , then open up a new frontier.

%%%%%%%%%%%%%%%%%%%%%%%%%%%%%%%%%%%%%%%%%%%%%%%%%%%%%%%%%%%%%%%
\section{Model and method}

We take, as the simplest model for correlated electrons, the single-band Hubbard model driven by an electric field with 
the Hamiltonian
\begin{align}
H(t)
%  &=
%    H_{\rm kin}(t)+H_{\rm int}
%  \nonumber
%  \\
  =
    \sum_{ij,\sigma} t_{ij}\exp\left(-i\int_{{\bm R}_j}^{{\bm R}_i}d{\bm r}\cdot {\bm A}(t)\right) c_{i\sigma}^\dagger c_{j\sigma}
    +H_{\rm int}(U),
\label{hamiltonian}
%\\
%&H_{\rm int}(U)
%  =
%    U\sum_i \left(n_{i\uparrow}-\frac{1}{2}\right)\left(n_{i\downarrow}-\frac{1}{2}\right),
%\nonumber
\end{align}
where $t_{ij}$ is the hopping between sites at ${\bm R}_i$ and ${\bm R}_j$, the electric field ${\bm F}(t)=-\partial {\bm A}(t)/\partial t$ is expressed in terms of 
the vector potential ${\bm A}(t)$, and $c^\dagger$ ($c$) creates (annihilates) an electron.  For the interaction we take 
the particle-hole symmetric form
\begin{align}
H_{\rm int}(U)
  &=
    U\sum_i \left(n_{i\uparrow}-\frac{1}{2}\right)\left(n_{i\downarrow}-\frac{1}{2}\right), 
\label{interaction term}
\end{align}
where $U(\ge\!\!\!0)$ is the repulsive Coulomb interaction with $n_{i\sigma}=c_{i\sigma}^\dagger c_{i\sigma}$.
%@and the origin of the energy is shifted (taking $(n_{i\sigma}-1/2)$ in place of $n_{i\sigma}$) for a later purpose.
We apply a pulsed wave at $t=0$, and switch off the field at $t=\tau$. % ($\tau$: duration of the pulse).
For the DMFT, we consider a hypercubic lattice with the Gaussian density of states 
$D(\epsilon)=\frac{1}{\sqrt{\pi}W}e^{-\epsilon^2/W^2}$, \cite{GeorgesKotliarKrauthRozenberg1996}
and apply the field in the diagonal direction with 
${\bm F}(t)=F(t)(1,1,\dots)$. The band is assumed to be half-filled.
Throughout the paper, we use the bandwidth $W$ as the unit of energy, 
%set $e=a=\hslash=1$, 
and take the initial temperature to be $T=0.1$.

\section{Results}

\subsection{Noninteracting system}
Let us start with the noninteracting system. We focus on 
%the density of occupied states defined by the lesser Green function,
the momentum distribution defined by
$f({\bm k},t) = -i\tilde{G}^<_{\bm k}(t,t)=-iG^<_{{\bm k}+{\bm A}(t)}(t,t)$,
where $G_{\bm k}^<(t,t')$ [$\tilde{G}^<_{\bm k}(t,t')$]
is the (gauge-invariant \cite{TsujiOkaAoki2008}) lesser Green function.
For the noninteracting system, the lesser Green function is given by
%$N(\omega,t)\equiv -\frac{i}{2\pi}\int d\bar{t} e^{i\omega \bar{t}}G^<(t+\bar{t}/2,t-\bar{t}/2)$. 
\begin{align}
G_{0{\bm k}}^<(t,t')
  &=
    if_0(\epsilon_{\bm k})\exp\left(-i\int_{t'}^{t} d\bar{t}\epsilon_{{\bm k}-{\bm A}(\bar{t})}\right),
\end{align}
where $f_0(\epsilon)=1/(e^{\epsilon/T}+1)$ is the Fermi distribution, 
and $\epsilon_{\bm k}$ the band dispersion. 
%If sufficient time has passed after the pulse excitation, it is analytically given by
After the pulse excitation ($t>\tau$), the momentum distribution becomes
%\begin{align}
$f({\bm k},t)=f_0(\epsilon_{{\bm k}-{\bm \phi}})$
%\end{align}
%\begin{align}
%N(\omega,t)
%  &=
%    \sum_{\bm k} f(\epsilon_{\bm k}) \delta(\omega-\epsilon_{{\bm k}-{\bm \phi}}),
%\label{occupation}
%\end{align}
%where $f(\epsilon)=1/(e^{\epsilon/T}+1)$ is the Fermi distribution function, 
%with $T$ the initial temperature, 
%$\epsilon_{\bm k}$ is the band dispersion, 
%$\mu$ is the chemical potential, 
with $\bm \phi=-{\bm A}(\tau)=\phi(1,1,\dots)$.
%with $\phi$ the dynamical phase (\ref{dynamical phase}).   
Note that the effect of the pulse field on the final state amounts to a momentum shift $\phi$ (\ref{dynamical phase}).
%When $\sin\phi\ll 1$, 
%(i.e., $\sin\phi\ll\cos\phi$), 
%we can explicitly evaluate Eq.~(\ref{occupation}) using the joint density of states 
%\cite{TurkowskiFreericks2005,TsujiOkaAoki2008} as
%\begin{align}
%N(\omega,t)
%  &=
%    \frac{1}{\pi}\int d\epsilon\int dv e^{-\epsilon^2-v^2}f(\epsilon)\delta(\omega-\epsilon\cos\phi+v\sin\phi)
%  \nonumber
%  \\
%  &=
%    \frac{1}{\pi|\cos\phi |}\int dv e^{-(v+\omega\sin\phi)^2/\cos^2\phi-\omega^2}
%    f\left(\frac{\omega+v\sin\phi}{\cos\phi}\right)
%  \nonumber
%  \\
%  &\sim
%   D(\omega)f(\omega\cos\phi).
%  \nonumber
%\end{align}
For a $\pi$ shift ($\phi\simeq\pi$), the electrons occupy the 
band top with $f({\bm k},t)\sim f_0(-\epsilon_{\bm k})$,
which is characterized by an effective temperature $T_{\rm eff}=-T<0$.
%\begin{align}
%T_{\rm eff}
%  &=
%   \frac{T}{\cos \phi},
%\end{align}
%which is negative for $\phi\sim\pi$. 
%@This may be viewed as a Bloch oscillation terminated at a finite $t$.
Thus a $\pi$ shift (which may be viewed as a partial Bloch oscillation)
is the condition that leads to a negative-$T$ state in the noninteracting system.

\subsection{Interacting system}
Now let us move on to the interacting case. 
There, we can identify the repulsion-to-attraction transition from 
the total energy $E_{\rm tot}(t)=\langle H(t)\rangle$: After the pulse excitation, a (nonintegrable) isolated system is supposed to 
approach a thermalized state \cite{RigolDunjkoOlshanii2008} with some effective temperature $T_{\rm eff}$ and a total energy $E_{\rm tot}(\tau)$ 
(which is conserved after the pulse is over at $t=\tau$). 
A thermal state with a positive temperature always gives $E_{\rm tot}<0$ at half filling 
for the interaction term (\ref{interaction term}),
%since the double occupancy $\langle n_{i\uparrow}n_{i\downarrow}\rangle<1/4$, 
while one with a negative temperature gives $E_{\rm tot}>0$.
This suggests that the total energy plays the role of an ``order parameter'' for the repulsion-to-attraction transition. If and only if $E_{\rm tot}(\tau)>0$
the system arrives at a negative-$T$ state ($T_{\rm eff}<0$), 
in which case the density matrix is given by
\begin{align}
\rho
  &\propto
    \exp\left(-\frac{1}{T_{\rm eff}}\left[\sum_{{\bm k},\sigma} \epsilon_{{\bm k}+{\bm \phi}}c_{{\bm k}\sigma}^\dagger c_{{\bm k}\sigma}
    +H_{\rm int}(U)\right]\right)
    %\sum_i \left(n_{i\uparrow}-\frac{1}{2}\right)\left(n_{i\downarrow}-\frac{1}{2}\right)\right]\right)
  \nonumber
  \\
  &=
    \exp\left(-\frac{1}{|T_{\rm eff}|}\left[\sum_{{\bm k},\sigma}\epsilon_{\bm k}\tilde{c}_{{\bm k}\sigma}^\dagger \tilde{c}_{{\bm k}\sigma}
    +\tilde{H}_{\rm int}(-U)\right]\right) .
    %\sum_i \left(\tilde{n}_{i\uparrow}-\frac{1}{2}\right)\left(\tilde{n}_{i\downarrow}-\frac{1}{2}\right)\right]\right).
  \label{density matrix}
\end{align}
Here we have introduced a gauge transformation 
\begin{align*}
c_{i\sigma} 
  &\to 
    \tilde{c}_{i\sigma}=e^{-i(\phi+\pi)(1,1,\dots)\cdot{\bm R}_i}c_{i\sigma}
\end{align*}
with $\tilde{H}_{\rm int}(-U)=-U\sum_i \left(\tilde{n}_{i\uparrow}-\frac{1}{2}\right)\left(\tilde{n}_{i\downarrow}-\frac{1}{2}\right)$ 
(with $\tilde{n}_{i\sigma}=\tilde{c}_{i\sigma}^\dagger \tilde{c}_{i\sigma}$) 
to cancel the phase shift in the kinetic energy.  
The above equation implies that the state can be viewed as a thermal state with a positive $T$ and 
an attractive interaction $-U<0$. This is the basic mechanism behind the repulsion-to-attraction transition driven by the pulse.  
Note that the condition for the interacting system is $T_{\rm eff}<0$, 
as opposed to the noninteracting counterpart $\phi\simeq\pi$.

\begin{figure}[t]
\includegraphics[width=8.5cm]{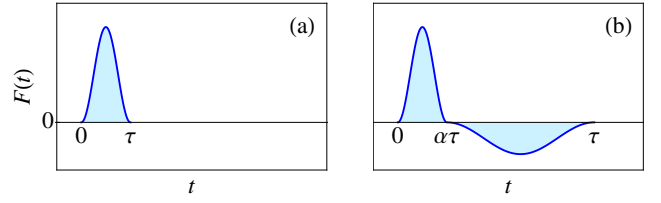}
\caption{(Color online) Schematic temporal profiles of  a half-cycle pulse (a) 
and  a monocycle pulse (b) for which $\alpha$ controls the asymmetry.}
\label{pulse shape}
\end{figure}
To make our argument more precise, we consider two types of pulses. 
One is a half-cycle pulse [Fig.~\ref{pulse shape}(a)], and the other is a monocycle pulse [Fig.~\ref{pulse shape}(b)]:
%The former is given by
\begin{align}
F_{\text{half-cycle}}(t)
  &=
    \frac{A}{\tau} s\bigg(\frac{t}{\tau}\bigg),
\label{half-cycle pulse}
%\end{align}
%while the latter is defined by
%\begin{align}
\\
F_{\text{monocycle}}(t)
  &=
    \frac{A}{\alpha\tau} s\bigg(\frac{t}{\alpha\tau}\bigg)
    -\frac{A}{(1-\alpha)\tau} s\bigg(\frac{\tau-t}{(1-\alpha)\tau}\bigg).
\label{monocycle pulse}
\end{align}
Here $A$ controls the amplitude of the pulse, $s(x)$ ($\geqslant\! 0$) is a pulse shape function that has support in $0\leqslant x\leqslant 1$ with 
$\int_0^1 dx s(x)=1$, and $\alpha$ ($0<\alpha<1$) controls 
the asymmetry of the monocycle pulse, with $\alpha=\frac{1}{2}$ corresponding to the symmetric case
[$F(\tau-t)=-F(t)$].
The dynamical phase (\ref{dynamical phase}) is $\phi=A$ ($\phi=0$)
for the half- (mono-)cycle pulse.

\begin{figure}[t]
\includegraphics[width=8.5cm]{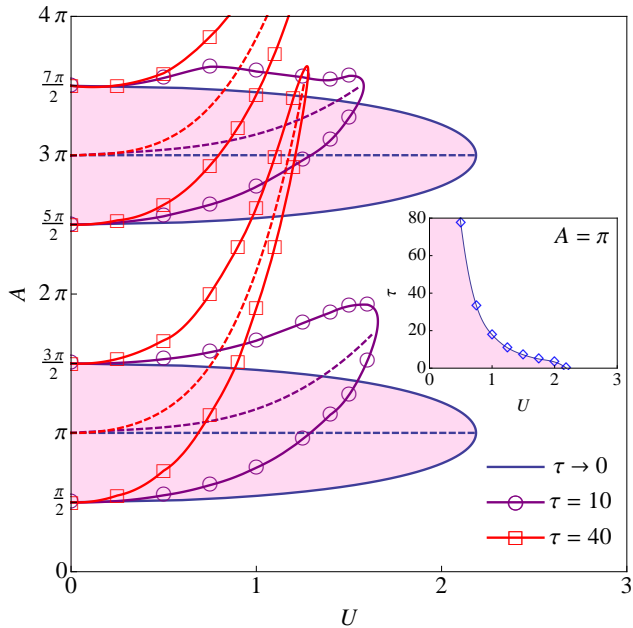}
\caption{(Color online) Phase diagram in the $(U,A)$ plane for the Hubbard model driven by the half-cycle cosine pulse.  
The regions surrounded by solid curves represent the pulse-induced, attractively interacting phase, while 
the dashed curves are loci of the shift $\delta = (2n+1)\pi$.
The inset shows the phase diagram in the $(U,\tau)$ plane for $A=\pi$.
%The colored region shows the attractively interacting phase.
}
\label{phase diagram}
\end{figure}
\subsection{Half-cycle pulse}
We first consider the half-cycle pulse (\ref{half-cycle pulse}). 
The simplest case is the limit $\tau\to 0$,
corresponding to a delta-function pulse [$F(t)\to A\delta(t)$].
In this case, the momentum shift is $\delta=\phi$, so that the order parameter reads
\begin{align*}
E_{\rm tot}(\tau)
  &=
    E_{\rm kin}(0)\cos\phi-iJ(0)\sin\phi+E_{\rm int}(0),
\end{align*}
where $E_{\rm kin}(t)$, $J(t)$, $E_{\rm int}(t)$ are the kinetic energy, current, and interaction energy at time $t$, respectively.
%the current $J=\langle i\sum_{{\bm k},\sigma} v_{\bm k} c_{{\bm k}\sigma}^\dagger c_{{\bm k}\sigma}\rangle=0$
%Since the system is initially in equilibrium, $J(0)=0$. 
Since $J(0)=0$ in the initial state and $\phi=A$ for the half-cycle pulse (\ref{half-cycle pulse}),  
we can identify the condition for the repulsion-to-attraction transition,
\begin{align}
E_{\rm kin}(0)\cos A+E_{\rm int}(0)>0, 
\label{condition}
\end{align}
which is completely determined by the equilibrium state.
The criterion is quite general, so should be applicable to systems
with any fillings in any dimensions
if one puts the origin of the energy to be the one at $T=\pm\infty$.
In Fig.~\ref{phase diagram}, we show the attractively interacting regions by the hashed areas.
Attractive regions appear periodically in the amplitude $A$ of the pulse  
as a series of lobes around $A\simeq (2n+1)\pi$ ($n=0,1,2,\dots$). 
%Since $E_{\rm kin}$ ($E_{\rm int}$) monotonically increases (decreases) with $U$, they cross at some point. In particular,
Each lobe has the tip at $U_{c}=2.186$, 
%i.e., the attractive region occurs for $U<U_c$.  Since $U_c$ 
which turns out to be smaller than the critical $U$ for the Mott transition, \cite{GeorgesKotliarKrauthRozenberg1996}
so that the transition always occurs in the metallic regime.   
The repulsion-to-attraction conversion is obviously distinct from a heating effect, since 
it appears and disappears 
repeatedly as one increases the amplitude $A$ of the pulse field. 

To study how the system evolves in time, we have numerically solved the model (\ref{hamiltonian}) with 
the nonequilibrium DMFT. \cite{FreericksTurkowskiZlatic2006}
%which takes into account the nonequilibrium correlation effects.
As an impurity solver for DMFT, we mainly employ the continuous-time quantum Monte Carlo method 
\cite{GullMillisLichtensteinRubtsovTroyerPhilipp2011} with the weak-coupling expansion generalized to nonequilibrium. \cite{WernerOkaMillis2009}
To capture the long-time ($t\geqslant 20$) behavior in the very weak-coupling 
regime ($U\leqslant 1.2$) we use 
the iterative perturbation theory, \cite{GeorgesKotliar1992,GeorgesKotliarKrauthRozenberg1996}
which is known to give quite accurate results up to a long time \cite{EcksteinWerner2011b} for such small $U$ at half filling. 
%If the method is not specified, it should be understood as the QMC solver.

\begin{figure}[t]
\begin{center}
\includegraphics[width=8.5cm]{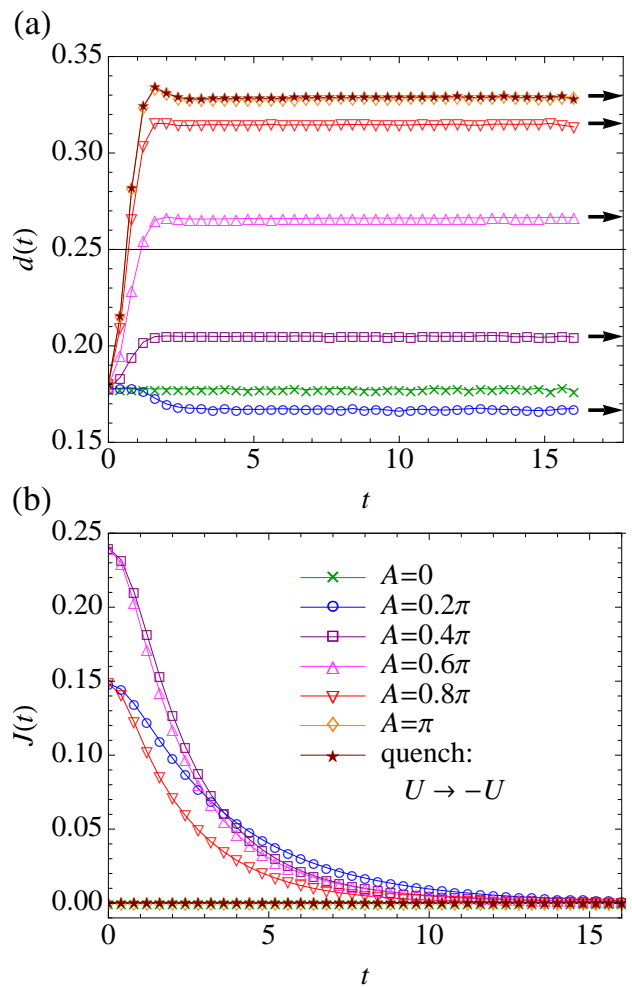}
\caption{(Color online) Time evolution of the double occupancy (a) and the current (b) after the delta-function-pulse excitation, which are compared with 
the interaction quench ($U\to -U$) with $U=1$.  Each arrow indicates the double occupancy in the corresponding thermal state 
with the same total energy.}
\label{delta-function pulse}
\end{center}
\end{figure}
%\begin{figure}[htbp]
%\begin{center}
%\includegraphics[width=8.5cm]{momentum-distribution.eps}
%\caption{The snapshots of the momentum distribution (black curves) at $t=0,0.8,1.6,\dots,16$ for the delta-function pulse
%with (a) $A=0.8\pi$ and (b) $A=\pi$, and $U=1$. They converge to thermal distributions with $T=-0.725$ and $T=-0.446$ (red curves), respectively.}
%\label{momentum distribution}
%\end{center}
%\end{figure}
In Fig.~\ref{delta-function pulse}(a) we show how the double occupancy $d(t) \equiv \langle n_{\uparrow}n_{\downarrow}(t)\rangle$, a measure of the interaction, 
%and the current $J(t)$ 
evolves after the delta-function pulse in an initially repulsive system ($U=1$).
%and converges quickly to thermal values (arrows in Fig.~\ref{delta-function pulse}) with the same total energies. 
We notice that for $A>0.5\pi$ $d(t)$ shoots 
well beyond the noninteracting value
$d=\langle n_{\uparrow} \rangle \langle n_{\downarrow} \rangle=0.25$, 
which implies that the electrons do indeed 
start to attract each other
after the pulse, as predicted from the criterion (\ref{condition}).
The repulsion-to-attraction transition is ``perfect" for $A=\pi$, 
where the temporal evolution of $d$ is 
found to accurately agree with that for the interaction quench, $U\to -U$ [Fig.~\ref{delta-function pulse}(a)].
%in Fig.~\ref{delta-function pulse}. 
For this ``$\pi$ pulse", the shift of the momentum just changes the sign of the hopping
($\epsilon_{\bm k}\to \epsilon_{{\bm k}+{\bm \phi}}=-\epsilon_{\bm k}$), which is known to be %@similar 
equivalent to interaction quench. \cite{TsujiOkaWernerAoki2011}

\begin{figure}[t]
\begin{center}
\includegraphics[width=8.5cm]{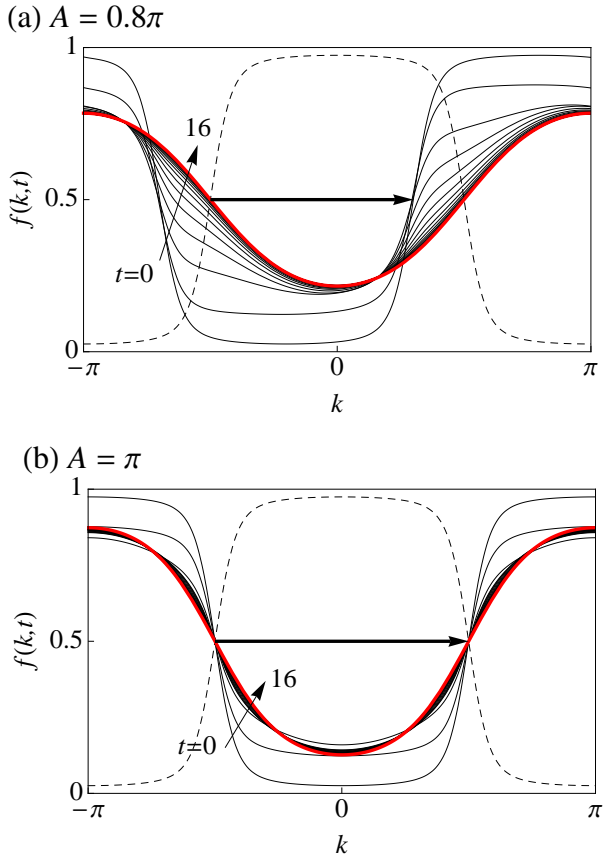}
\caption{(Color online) Snapshots of the corresponding momentum distribution $f(k,t)$ \cite{MomentumDistribution} (black curves) at $t=0,0.8,1.6,\dots,16$
are shown for $A=0.8\pi$ (a) and  $A=\pi$ (b).  They converge to thermal distributions with $T_{\rm eff}=-0.725$ and $T_{\rm eff}=-0.446$ (thick red curves), respectively.
Dashed curves represent the initial distributions. The horizontal arrows indicate the pulse-induced phase shift $\delta=\phi=A$.}
\label{momentum snapshot}
\end{center}
\end{figure}
Remarkably, after the pulse excitation $d(t)$ relaxes quickly to a steady state ($t\lesssim 3$).
%This is understood as a thermalization: Since the total energy is conserved after the pulse excitation, we can extract
%a thermal state having the same total energy.
We have confirmed that it converges to the thermal value $d_{\rm th}$ 
[indicated by arrows in Fig.~\ref{delta-function pulse}(a)] for the equilibrium state having the same $E_{\rm tot}$.
For $A>0.5\pi$ the corresponding temperature ($T_{\rm eff}$) of the thermal state is negative since $E_{\rm tot}>0$.
Note that $d_{\rm th}$ is a nonmonotonic function of temperature, so that $d(t)$ decreases in time for $A=0.2\pi$.

On the other hand, the current $J(t)$ [Fig.~\ref{delta-function pulse}(b)] generated by the momentum shift $k\to k+\phi$
decays more slowly ($t\lesssim 15$) for $0<A<\pi$ than $d(t)$. %the double occupancy. 
%the current is generated immediately after the pulse excitation. 
The slow relaxation is also seen in the momentum distributions $f(k,t)$ \cite{MomentumDistribution} 
[Fig.~\ref{momentum snapshot}(a) for $A=0.8\pi$ and (b) for $A=\pi$].
%To understand the difference between the two relaxation times, we plot the momentum distribution $f(k,t)$ \cite{MomentumDistribution}
%at various $t$ for $A=0.8\pi$ [Fig.~\ref{delta-function pulse}(c)] and $A=\pi$ [Fig.~\ref{delta-function pulse}(d)].
%in Fig.~\ref{momentum distribution}. 
A similar separation of the relaxation times of $d(t)$ and $f(k,t)$ has been observed in the interaction quench, \cite{EcksteinKollarWerner2009}
and was attributed to the existence of a ``prethermalized" state. \cite{MoeckelKehrein2008}
Here the slow decay becomes particularly evident when the momentum shift is not perfect (i.e., $A\neq \pi$).
%In both cases, the jump height at the Fermi edge in the initially shifted distribution decreases rapidly,
%and the distribution for $A=\pi$ almost arrives at a thermalized state at $t=2.4$. If $A=0.8\pi$, however, 
In this case the system needs to adjust the momentum shift to $\pi$ to achieve a thermal state.
%In order to arrive at the thermalized state with the momentum shift 0 ($T_{\rm eff}>0$) or $\pi$ ($T_{\rm eff}<0$), 
Since the relaxation involves a violation of the momentum 
conservation by Umklapp scattering, it takes longer than the relaxation of the double occupancy via particle-hole annihilations. 
%The separation of the time scale of two relaxation processes is clearly seen in the momentum distribution $f(\kappa)$ (Fig.~\ref{momentum distribution}).
The distributions eventually 
relax to thermal states with $T_{\rm eff}<0$ [red curves in Fig.~\ref{momentum snapshot}(a), (b)]. 
%, which demonstrate the negative-$T$ thermalization after the pulse drive.

%\begin{figure}[htbp]
%\includegraphics[width=8.5cm]{phase-diagram-U-vs-tau.eps}
%\caption{A phase diagram of the Hubbard model driven by the half-cycle cosine pulse with $A=\pi$.
%The colored region shows the attractively interacting phase.}
%\label{U vs tau}
%\end{figure}
So far we have examined the delta-function pulse ($\tau\to 0$). 
To be more realistic it is important to evaluate the effect of $\tau$ on the transition. Here we take, as an example,
a half-cycle pulse with 
\begin{align}
s(x)
  &=
    1-\cos(2\pi x), 
\label{cosine pulse}
\end{align}
which we call the ``cosine pulse".  
We can again use $E_{\rm tot}(\tau)$ as an indicator for the interaction conversion.
%whether the interaction is effectively converted to an attraction.
In the inset of Fig.~\ref{phase diagram}, we show how the critical interaction ($U_c$) of the repulsion-to-attraction transition induced by the cosine pulse
with $A=\pi$ depends on $\tau$. For $\tau\lesssim 10$, $U_c(\tau)$ rapidly falls off from $U_{c}(\tau=0)$,
while for larger $\tau$ it decays to zero very slowly. 
%The behavior is roughly understood as a crossover from a nonadiabatic to adiabatic process. 
In the adiabatic limit ($\tau\to\infty$) 
%the momentum shift ($\epsilon_{\bm k}\to\epsilon_{{\bm k}-{\bm \phi}}$)
%, and it remains in the initial state. 
%The bigger $\tau$ is, the more adiabatic the ramp up/down process of the field becomes.
%the total energy does not change so much from the initial value.
$E_{\rm tot}(\tau) \to E_{\rm tot}(0)<0$ for $U>0$ and the repulsion-to-attraction transition naturally disappears.
%Note that the delta function can be treated exactly in the numerical calculation because the vector potential
%always appears as an integrand $\int_{t'}^{t} d\bar{t} F(\bar{t})$ which is a non-singular function of time.

\begin{figure}[t]
\includegraphics[width=8.5cm]{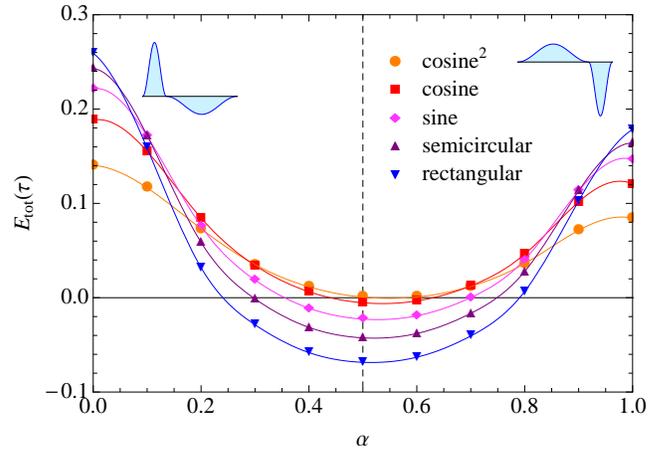}
\caption{(Color online) The total energies at $t=\tau$ for the system with $U=1$ driven by monocycle pulses
with $A=\pi$, $\tau=40$ plotted for various pulse shapes as a function of the asymmetry parameter $\alpha$.}
\label{asymmetry}
\end{figure}
For general $A$ and non-zero $\tau$ of the cosine pulse, Fig.~\ref{phase diagram} depicts the phase diagram for the pulse-driven Hubbard model.  
The attractive regions now deform in a characteristic 
manner, i.e., the tips of the lobes bend toward larger $A$, 
which becomes drastic for $\tau=40$.
The deformation of the phase diagram implies a rather counterintuitive fact: 
for $\tau=40$ and $U\sim 0.9$ 
%around the region where the deformed lobe traverses the next ($3\pi$-pulse) phase (e.g., around $\tau=40, U\sim 1$) 
the repulsion-to-attraction conversion occurs even for a trivial phase $\phi=A=2\pi$.
%which would  not give an attractive interaction per se, 
%the conversion to attraction does occur. 
This suggests that the effective phase shift $\delta$ that the correlated system acquires is not equal to $\phi$ for nonzero $\tau$.  
%We can In fact confirm that  
%the @extrema? 
%ridge line --- meaning? 
Using the extremal points 
of the total energy satisfying $\partial E_{\rm tot}(\tau)/\partial A=0$ 
(dashed curves in Fig.~\ref{phase diagram}) as an estimate for $\delta=(2n+1)\pi$, we can see the large deviations of $\delta$
from $\phi=A=(2n+1)\pi$ as $U$ and $\tau$ grow.  We attribute this to a 
correlation effect: During irradiation with the pulse
the electrons scatter with each other, which causes broadening of the momentum distribution.
Consequently the shift in the momentum is suppressed, and $\delta$ becomes smaller than $\phi$.

\subsection{Monocycle pulse}
The result that $\delta\neq\phi$
%, along with 
%the momentum shift strongly dependent on the pulse duration, 
%lead us to predict 
suggests an experimentally much simpler way to induce the repulsion-to-attraction transition ($\delta\simeq\pi$) 
by monocycle pulses (\ref{monocycle pulse}) with $\phi=0$.  
The basic idea is the following: since the effect of a half-cycle pulse very much depends on its width, 
we can suitably choose the widths of the first and second half cycles of a monocycle pulse
so that the total phase shift is $\simeq\pi$.
%a monocycle pulse with suitably chosen widths
%for the first half and second half cycles should be able to  
%induce the negative-$T$ state.  
In Fig.~\ref{asymmetry}, we plot the total energy at
$t=\tau$ for various types of pulse shapes. The shape function of each pulse is defined by
\begin{align*}
s(x)
  &=
    \begin{cases}
    \frac{2}{3}[1-\cos(2\pi x)]^2 & \text{cosine$^2$  pulse}, \\
    \frac{\pi}{2}\sin(\pi x) & \text{sine pulse}, \\
    \frac{4}{\pi} \sqrt{1-(2x-1)^2} & \text{semicircular pulse}, \\
    1 & \text{rectangular pulse},
    \end{cases}
\end{align*}
for $0\leqslant x\leqslant 1$. $s(x)$ of the cosine pulse is defined as before [Eq.~(\ref{cosine pulse})]. 
%and $s(x)=\delta(x-1/2)$ (delta-function pulse).
From the above argument, the asymmetry, here represented by 
$\alpha$, should be important, and we can 
indeed see that $E_{\rm tot}(\tau)$ becomes positive 
(implying a repulsion-to-attraction transition)  
as soon as we go sufficiently away from the symmetric pulse form ($\alpha=\frac{1}{2}$).  To be more precise, 
the momentum shift induced by the first half cycle does not cancel
the one induced by the second half cycle when the mono-pulse is sufficiently asymmetric.
%Surprisingly, the cosine$^2$ pulse gives $E_{\rm tot}(\tau)>0$ even at $\alpha=\frac{1}{2}$, since the system becomes less sensitive to the second half cycle
Note that the cosine$^2$ pulse gives $E_{\rm tot}(\tau)>0$ even at $\alpha=\frac{1}{2}$, although its value is very small. 
%since the system becomes less sensitive to the second half cycle after the excitation. 
%@This means that the asymmetry is not a necessary condition for the repulsion-to-attraction conversion, but it makes it easier to induce the transition. --- realistically it is a necessary condition I think
There is a slight difference in $E_{\rm tot}(\tau)$ for the cases with 
$\alpha<\frac{1}{2}$ (i.e., the sharp pulse comes first, followed by the broad one) and $\alpha>\frac{1}{2}$, 
implying that the former is more suitable than the latter to induce the attractive interaction.
The order parameter $E_{\rm tot}(\tau)$ also depends on the shape of the pulse. In the case of $U=1$ and $\tau=40$,
the larger the peak amplitude [$s\left(\frac{1}{2}\right)=\frac{8}{3},2, \frac{\pi}{2}, \frac{4}{\pi}, 1$ for the cosine$^2$, cosine, sine, semicircular, rectangular pulse, respectively], 
the larger the $E_{\rm tot}(\tau)$ around $\alpha=\frac{1}{2}$.  
For even shaper pulse shapes (cosine$^3$, \ldots), $E_{\rm tot}(\tau)$ at $\alpha=\frac{1}{2}$ starts to decrease, so that the cosine$^2$ pulse is an optimal shape in this case.
%For the cosine pulse, for example, an asymmetry of $\alpha=0.4$ is sufficient to induce 
%the repulsion-to-attraction transition. Such pulses can be easily reproduced with current laser techniques \cite{JonesYouBucksbaum1993,Hebling2008,JawariyaNagaiTanaka2009}.

\section{Discussion and outlook}

Finally, let us discuss the experimental feasibility of the pulse-induced 
repulsion-to-attraction transition proposed here.  
Asymmetric monocycle pulses with qualitative features comparable to the shapes considered here can be generated 
experimentally. \cite{Hauri}
%\footnote{C. Hauri, private discussions.}
%.... @examples and discussions with Shimano, Hauri. we can even cite experimental papers.
One way to detect the negative-$T$ state is 
to measure the time-resolved dc or optical conductivity, which will become negative after the pulse irradiation due to energy gain.
%if the negative-$T$ state is realized. 
Another possibility is to measure momentum-resolved photoemission spectra, which can
detect the shift in the momentum distribution [Fig.~\ref{momentum snapshot}(a), (b)]. 
We require the time resolution of the measurement to be fine enough that it can detect
the population-inverted state before it relaxes to a more stable state through energy dissipation.
The dissipation typically occurs due to phonons whose time scale is of the order of 0.1$-$1 ps, \cite{Perfetti2007}
which allows one to access the negative-$T$ state using current ultrafast laser techniques with a resolution $\sim$ 10 fs. \cite{Wall2011}
%Measurements of quantities related to charge and/or spin fluctuations are also useful since the switching of the interaction from repulsive to attractive
%effectively interchanges the charge and spin degrees of freedom. 
Materials that have a metallic band at the Fermi energy, well  
separated from the other bands, are suitable candidates because of the absence of interband transitions
that destabilize the population-inverted state. As an example, transparent conductors (e.g., Sn-doped In$_2$O$_3$ \cite{MryasovFreeman2001}) and
alkali-metal-loaded zeolites \cite{Arita2004} such as sodalite \cite{NakamuraKoretsuneArita2009}
%@ R. Arita et al, Phys. Rev. B {\bf 69}, 195106 (2004).
are materials that exhibit this kind of band structure.

\section{Acknowledgments}

We thank C. Hauri, C. Mudry and R. Shimano for stimulating discussions and S. Biermann for suggesting Sn-doped In$_2$O$_3$ as a potentially interesting material. 
This work was supported by SNF Grant No. PP0022-118866, and by a Grant-in-Aid for Scientific Research on Innovative Areas 
``Optical Science of Dynamically Correlated Electrons''.

\bibliographystyle{apsrev}
\bibliography{pulse}

\begin{thebibliography}{35}
\expandafter\ifx\csname natexlab\endcsname\relax\def\natexlab#1{#1}\fi
\expandafter\ifx\csname bibnamefont\endcsname\relax
  \def\bibnamefont#1{#1}\fi
\expandafter\ifx\csname bibfnamefont\endcsname\relax
  \def\bibfnamefont#1{#1}\fi
\expandafter\ifx\csname citenamefont\endcsname\relax
  \def\citenamefont#1{#1}\fi
\expandafter\ifx\csname url\endcsname\relax
  \def\url#1{\texttt{#1}}\fi
\expandafter\ifx\csname urlprefix\endcsname\relax\def\urlprefix{URL }\fi
\providecommand{\bibinfo}[2]{#2}
\providecommand{\eprint}[2][]{\url{#2}}

\bibitem[{\citenamefont{Micnas et~al.}(1990)\citenamefont{Micnas, Ranninger,
  and Robaszkiewicz}}]{MicnasRanningerRobaszkiewicz1990}
\bibinfo{author}{\bibfnamefont{R.}~\bibnamefont{Micnas}},
  \bibinfo{author}{\bibfnamefont{J.}~\bibnamefont{Ranninger}},
  \bibnamefont{and}
  \bibinfo{author}{\bibfnamefont{S.}~\bibnamefont{Robaszkiewicz}},
  \bibinfo{journal}{Rev. Mod. Phys.} \textbf{\bibinfo{volume}{62}},
  \bibinfo{pages}{113} (\bibinfo{year}{1990}).

\bibitem[{\citenamefont{Keller et~al.}(2001)\citenamefont{Keller, Metzner, and
  Schollw\"ock}}]{KellerMetznerSchollwoeck2001}
\bibinfo{author}{\bibfnamefont{M.}~\bibnamefont{Keller}},
  \bibinfo{author}{\bibfnamefont{W.}~\bibnamefont{Metzner}}, \bibnamefont{and}
  \bibinfo{author}{\bibfnamefont{U.}~\bibnamefont{Schollw\"ock}},
  \bibinfo{journal}{Phys. Rev. Lett.} \textbf{\bibinfo{volume}{86}},
  \bibinfo{pages}{4612} (\bibinfo{year}{2001}).

\bibitem[{\citenamefont{Nozi$\grave{\rm e}$res and
  Schmitt-Rink}(1985)}]{NozieresSchmittRink1985}
\bibinfo{author}{\bibfnamefont{P.}~\bibnamefont{Nozi$\grave{\rm e}$res}}
  \bibnamefont{and}
  \bibinfo{author}{\bibfnamefont{S.}~\bibnamefont{Schmitt-Rink}},
  \bibinfo{journal}{J. Low Temp. Phys.} \textbf{\bibinfo{volume}{59}},
  \bibinfo{pages}{195} (\bibinfo{year}{1985}).

\bibitem[{\citenamefont{Manmana et~al.}(2007)\citenamefont{Manmana, Wessel,
  Noack, and Muramatsu}}]{ManmanaWesselNoackMuramatsu2007}
\bibinfo{author}{\bibfnamefont{S.~R.} \bibnamefont{Manmana}},
  \bibinfo{author}{\bibfnamefont{S.}~\bibnamefont{Wessel}},
  \bibinfo{author}{\bibfnamefont{R.~M.} \bibnamefont{Noack}}, \bibnamefont{and}
  \bibinfo{author}{\bibfnamefont{A.}~\bibnamefont{Muramatsu}},
  \bibinfo{journal}{Phys. Rev. Lett.} \textbf{\bibinfo{volume}{98}},
  \bibinfo{pages}{210405} (\bibinfo{year}{2007}).

\bibitem[{\citenamefont{Moeckel and Kehrein}(2008)}]{MoeckelKehrein2008}
\bibinfo{author}{\bibfnamefont{M.}~\bibnamefont{Moeckel}} \bibnamefont{and}
  \bibinfo{author}{\bibfnamefont{S.}~\bibnamefont{Kehrein}},
  \bibinfo{journal}{Phys. Rev. Lett.} \textbf{\bibinfo{volume}{100}},
  \bibinfo{pages}{175702} (\bibinfo{year}{2008}).

\bibitem[{\citenamefont{Eckstein et~al.}(2009)\citenamefont{Eckstein, Kollar,
  and Werner}}]{EcksteinKollarWerner2009}
\bibinfo{author}{\bibfnamefont{M.}~\bibnamefont{Eckstein}},
  \bibinfo{author}{\bibfnamefont{M.}~\bibnamefont{Kollar}}, \bibnamefont{and}
  \bibinfo{author}{\bibfnamefont{P.}~\bibnamefont{Werner}},
  \bibinfo{journal}{Phys. Rev. Lett.} \textbf{\bibinfo{volume}{103}},
  \bibinfo{pages}{056403} (\bibinfo{year}{2009}).

\bibitem[{\citenamefont{Bloch et~al.}(2008)\citenamefont{Bloch, Dalibard, and
  Zwerger}}]{BlochDalibardZwerger2008}
\bibinfo{author}{\bibfnamefont{I.}~\bibnamefont{Bloch}},
  \bibinfo{author}{\bibfnamefont{J.}~\bibnamefont{Dalibard}}, \bibnamefont{and}
  \bibinfo{author}{\bibfnamefont{W.}~\bibnamefont{Zwerger}},
  \bibinfo{journal}{Rev. Mod. Phys.} \textbf{\bibinfo{volume}{80}},
  \bibinfo{pages}{885} (\bibinfo{year}{2008}).

\bibitem[{\citenamefont{Chin et~al.}(2010)\citenamefont{Chin, Grimm, Julienne,
  and Tiesinga}}]{ChinGrimmJulienneTiesinga2010}
\bibinfo{author}{\bibfnamefont{C.}~\bibnamefont{Chin}},
  \bibinfo{author}{\bibfnamefont{R.}~\bibnamefont{Grimm}},
  \bibinfo{author}{\bibfnamefont{P.}~\bibnamefont{Julienne}}, \bibnamefont{and}
  \bibinfo{author}{\bibfnamefont{E.}~\bibnamefont{Tiesinga}},
  \bibinfo{journal}{Rev. Mod. Phys.} \textbf{\bibinfo{volume}{82}},
  \bibinfo{pages}{1225} (\bibinfo{year}{2010}).

\bibitem[{\citenamefont{Purcell and Pound}(1951)}]{PurcellPound1951}
\bibinfo{author}{\bibfnamefont{E.~M.} \bibnamefont{Purcell}} \bibnamefont{and}
  \bibinfo{author}{\bibfnamefont{R.~V.} \bibnamefont{Pound}},
  \bibinfo{journal}{Phys. Rev.} \textbf{\bibinfo{volume}{81}},
  \bibinfo{pages}{279} (\bibinfo{year}{1951}).

\bibitem[{\citenamefont{Ramsey}(1956)}]{Ramsey1956}
\bibinfo{author}{\bibfnamefont{N.~F.} \bibnamefont{Ramsey}},
  \bibinfo{journal}{Phys. Rev.} \textbf{\bibinfo{volume}{103}},
  \bibinfo{pages}{20} (\bibinfo{year}{1956}).

\bibitem[{\citenamefont{Rapp et~al.}(2010)\citenamefont{Rapp, Mandt, and
  Rosch}}]{RappMandtRosch2010}
\bibinfo{author}{\bibfnamefont{A.}~\bibnamefont{Rapp}},
  \bibinfo{author}{\bibfnamefont{S.}~\bibnamefont{Mandt}}, \bibnamefont{and}
  \bibinfo{author}{\bibfnamefont{A.}~\bibnamefont{Rosch}},
  \bibinfo{journal}{Phys. Rev. Lett.} \textbf{\bibinfo{volume}{105}},
  \bibinfo{pages}{220405} (\bibinfo{year}{2010}).

\bibitem[{\citenamefont{Tsuji et~al.}(2011)\citenamefont{Tsuji, Oka, Werner,
  and Aoki}}]{TsujiOkaWernerAoki2011}
\bibinfo{author}{\bibfnamefont{N.}~\bibnamefont{Tsuji}},
  \bibinfo{author}{\bibfnamefont{T.}~\bibnamefont{Oka}},
  \bibinfo{author}{\bibfnamefont{P.}~\bibnamefont{Werner}}, \bibnamefont{and}
  \bibinfo{author}{\bibfnamefont{H.}~\bibnamefont{Aoki}},
  \bibinfo{journal}{Phys. Rev. Lett.} \textbf{\bibinfo{volume}{106}},
  \bibinfo{pages}{236401} (\bibinfo{year}{2011}).

\bibitem[{\citenamefont{Cavalieri et~al.}(2007)\citenamefont{Cavalieri, Muller,
  Uphues, Yakovlev, Baltuska, Horvath, Schmidt, Blumel, Holzwarth, Hendel
  et~al.}}]{Cavalieri2007}
\bibinfo{author}{\bibfnamefont{A.~L.} \bibnamefont{Cavalieri}},
  \bibinfo{author}{\bibfnamefont{N.}~\bibnamefont{Muller}},
  \bibinfo{author}{\bibfnamefont{T.}~\bibnamefont{Uphues}},
  \bibinfo{author}{\bibfnamefont{V.~S.} \bibnamefont{Yakovlev}},
  \bibinfo{author}{\bibfnamefont{A.}~\bibnamefont{Baltuska}},
  \bibinfo{author}{\bibfnamefont{B.}~\bibnamefont{Horvath}},
  \bibinfo{author}{\bibfnamefont{B.}~\bibnamefont{Schmidt}},
  \bibinfo{author}{\bibfnamefont{L.}~\bibnamefont{Blumel}},
  \bibinfo{author}{\bibfnamefont{R.}~\bibnamefont{Holzwarth}},
  \bibinfo{author}{\bibfnamefont{S.}~\bibnamefont{Hendel}},
  \bibnamefont{et~al.}, \bibinfo{journal}{Nature (London)}
  \textbf{\bibinfo{volume}{449}}, \bibinfo{pages}{1029} (\bibinfo{year}{2007}).

\bibitem[{\citenamefont{Wall et~al.}(2011)\citenamefont{Wall, Brida, Clark,
  Ehrke, Jaksch, Ardavan, Bonora, Uemura, Takahashi, Hasegawa
  et~al.}}]{Wall2011}
\bibinfo{author}{\bibfnamefont{S.}~\bibnamefont{Wall}},
  \bibinfo{author}{\bibfnamefont{D.}~\bibnamefont{Brida}},
  \bibinfo{author}{\bibfnamefont{S.~R.} \bibnamefont{Clark}},
  \bibinfo{author}{\bibfnamefont{H.~P.} \bibnamefont{Ehrke}},
  \bibinfo{author}{\bibfnamefont{D.}~\bibnamefont{Jaksch}},
  \bibinfo{author}{\bibfnamefont{A.}~\bibnamefont{Ardavan}},
  \bibinfo{author}{\bibfnamefont{S.}~\bibnamefont{Bonora}},
  \bibinfo{author}{\bibfnamefont{H.}~\bibnamefont{Uemura}},
  \bibinfo{author}{\bibfnamefont{Y.}~\bibnamefont{Takahashi}},
  \bibinfo{author}{\bibfnamefont{T.}~\bibnamefont{Hasegawa}},
  \bibnamefont{et~al.}, \bibinfo{journal}{Nat. Phys.}
  \textbf{\bibinfo{volume}{7}}, \bibinfo{pages}{114} (\bibinfo{year}{2011}).

\bibitem[{\citenamefont{Ulbricht et~al.}(2011)\citenamefont{Ulbricht, Hendry,
  Shan, Heinz, and Bonn}}]{UlbrichtHendryShanHeinz2011}
\bibinfo{author}{\bibfnamefont{R.}~\bibnamefont{Ulbricht}},
  \bibinfo{author}{\bibfnamefont{E.}~\bibnamefont{Hendry}},
  \bibinfo{author}{\bibfnamefont{J.}~\bibnamefont{Shan}},
  \bibinfo{author}{\bibfnamefont{T.~F.} \bibnamefont{Heinz}}, \bibnamefont{and}
  \bibinfo{author}{\bibfnamefont{M.}~\bibnamefont{Bonn}},
  \bibinfo{journal}{Rev. Mod. Phys.} \textbf{\bibinfo{volume}{83}},
  \bibinfo{pages}{543} (\bibinfo{year}{2011}).

\bibitem[{\citenamefont{Georges et~al.}(1996)\citenamefont{Georges, Kotliar,
  Krauth, and Rozenberg}}]{GeorgesKotliarKrauthRozenberg1996}
\bibinfo{author}{\bibfnamefont{A.}~\bibnamefont{Georges}},
  \bibinfo{author}{\bibfnamefont{G.}~\bibnamefont{Kotliar}},
  \bibinfo{author}{\bibfnamefont{W.}~\bibnamefont{Krauth}}, \bibnamefont{and}
  \bibinfo{author}{\bibfnamefont{M.~J.} \bibnamefont{Rozenberg}},
  \bibinfo{journal}{Rev. Mod. Phys.} \textbf{\bibinfo{volume}{68}},
  \bibinfo{pages}{13} (\bibinfo{year}{1996}).

\bibitem[{\citenamefont{Freericks et~al.}(2006)\citenamefont{Freericks,
  Turkowski, and Zlati\ifmmode~\acute{c}\else
  \'{c}\fi{}}}]{FreericksTurkowskiZlatic2006}
\bibinfo{author}{\bibfnamefont{J.~K.} \bibnamefont{Freericks}},
  \bibinfo{author}{\bibfnamefont{V.~M.} \bibnamefont{Turkowski}},
  \bibnamefont{and}
  \bibinfo{author}{\bibfnamefont{V.}~\bibnamefont{Zlati\ifmmode~\acute{c}\else
  \'{c}\fi{}}}, \bibinfo{journal}{Phys. Rev. Lett.}
  \textbf{\bibinfo{volume}{97}}, \bibinfo{pages}{266408}
  (\bibinfo{year}{2006}).

\bibitem[{\citenamefont{Jones et~al.}(1993)\citenamefont{Jones, You, and
  Bucksbaum}}]{JonesYouBucksbaum1993}
\bibinfo{author}{\bibfnamefont{R.~R.} \bibnamefont{Jones}},
  \bibinfo{author}{\bibfnamefont{D.}~\bibnamefont{You}}, \bibnamefont{and}
  \bibinfo{author}{\bibfnamefont{P.~H.} \bibnamefont{Bucksbaum}},
  \bibinfo{journal}{Phys. Rev. Lett.} \textbf{\bibinfo{volume}{70}},
  \bibinfo{pages}{1236} (\bibinfo{year}{1993}).

\bibitem[{\citenamefont{Hebling et~al.}(2008)\citenamefont{Hebling, Yeh,
  Hoffmann, Bartal, and Nelson}}]{Hebling2008}
\bibinfo{author}{\bibfnamefont{J.}~\bibnamefont{Hebling}},
  \bibinfo{author}{\bibfnamefont{K.-L.} \bibnamefont{Yeh}},
  \bibinfo{author}{\bibfnamefont{M.~C.} \bibnamefont{Hoffmann}},
  \bibinfo{author}{\bibfnamefont{B.}~\bibnamefont{Bartal}}, \bibnamefont{and}
  \bibinfo{author}{\bibfnamefont{K.~A.} \bibnamefont{Nelson}},
  \bibinfo{journal}{J. Opt. Soc. Am. B} \textbf{\bibinfo{volume}{25}},
  \bibinfo{pages}{B6} (\bibinfo{year}{2008}).

\bibitem[{\citenamefont{Jewariya et~al.}(2009)\citenamefont{Jewariya, Nagai,
  and Tanaka}}]{JawariyaNagaiTanaka2009}
\bibinfo{author}{\bibfnamefont{M.}~\bibnamefont{Jewariya}},
  \bibinfo{author}{\bibfnamefont{M.}~\bibnamefont{Nagai}}, \bibnamefont{and}
  \bibinfo{author}{\bibfnamefont{K.}~\bibnamefont{Tanaka}},
  \bibinfo{journal}{J. Opt. Soc. Am. B} \textbf{\bibinfo{volume}{26}},
  \bibinfo{pages}{A101} (\bibinfo{year}{2009}).

\bibitem[{\citenamefont{Moritz et~al.}(2010)\citenamefont{Moritz, Devereaux,
  and Freericks}}]{MoritzDevereauxFreericks2010}
\bibinfo{author}{\bibfnamefont{B.}~\bibnamefont{Moritz}},
  \bibinfo{author}{\bibfnamefont{T.~P.} \bibnamefont{Devereaux}},
  \bibnamefont{and} \bibinfo{author}{\bibfnamefont{J.~K.}
  \bibnamefont{Freericks}}, \bibinfo{journal}{Phys. Rev. B}
  \textbf{\bibinfo{volume}{81}}, \bibinfo{pages}{165112}
  (\bibinfo{year}{2010}).

\bibitem[{\citenamefont{Eckstein and
  Werner}(2011{\natexlab{a}})}]{EcksteinWerner2011a}
\bibinfo{author}{\bibfnamefont{M.}~\bibnamefont{Eckstein}} \bibnamefont{and}
  \bibinfo{author}{\bibfnamefont{P.}~\bibnamefont{Werner}},
  \bibinfo{journal}{Phys. Rev. B} \textbf{\bibinfo{volume}{84}},
  \bibinfo{pages}{035122} (\bibinfo{year}{2011}{\natexlab{a}}).

\bibitem[{hal()}]{half-cycle}
\bibinfo{note}{It is nevertheless possible to produce pulses containing only a
  half-cycle oscillation that gives nonzero $\phi$. See, e.g.,
  Ref.~\onlinecite{JonesYouBucksbaum1993}.}

\bibitem[{\citenamefont{Tsuji et~al.}(2008)\citenamefont{Tsuji, Oka, and
  Aoki}}]{TsujiOkaAoki2008}
\bibinfo{author}{\bibfnamefont{N.}~\bibnamefont{Tsuji}},
  \bibinfo{author}{\bibfnamefont{T.}~\bibnamefont{Oka}}, \bibnamefont{and}
  \bibinfo{author}{\bibfnamefont{H.}~\bibnamefont{Aoki}},
  \bibinfo{journal}{Phys. Rev. B} \textbf{\bibinfo{volume}{78}},
  \bibinfo{pages}{235124} (\bibinfo{year}{2008}).

\bibitem[{Rig()}]{RigolDunjkoOlshanii2008}
\bibinfo{note}{See, e.g., M. Rigol, V. Dunjko, and M. Olshanii, Nature (London)
  {\bf 452}, 854 (2008), and references therein.}

\bibitem[{\citenamefont{Gull et~al.}(2011)\citenamefont{Gull, Millis,
  Lichtenstein, Rubtsov, Troyer, and
  Werner}}]{GullMillisLichtensteinRubtsovTroyerPhilipp2011}
\bibinfo{author}{\bibfnamefont{E.}~\bibnamefont{Gull}},
  \bibinfo{author}{\bibfnamefont{A.~J.} \bibnamefont{Millis}},
  \bibinfo{author}{\bibfnamefont{A.~I.} \bibnamefont{Lichtenstein}},
  \bibinfo{author}{\bibfnamefont{A.~N.} \bibnamefont{Rubtsov}},
  \bibinfo{author}{\bibfnamefont{M.}~\bibnamefont{Troyer}}, \bibnamefont{and}
  \bibinfo{author}{\bibfnamefont{P.}~\bibnamefont{Werner}},
  \bibinfo{journal}{Rev. Mod. Phys.} \textbf{\bibinfo{volume}{83}},
  \bibinfo{pages}{349} (\bibinfo{year}{2011}).

\bibitem[{\citenamefont{Werner et~al.}(2009)\citenamefont{Werner, Oka, and
  Millis}}]{WernerOkaMillis2009}
\bibinfo{author}{\bibfnamefont{P.}~\bibnamefont{Werner}},
  \bibinfo{author}{\bibfnamefont{T.}~\bibnamefont{Oka}}, \bibnamefont{and}
  \bibinfo{author}{\bibfnamefont{A.~J.} \bibnamefont{Millis}},
  \bibinfo{journal}{Phys. Rev. B} \textbf{\bibinfo{volume}{79}},
  \bibinfo{pages}{035320} (\bibinfo{year}{2009}).

\bibitem[{\citenamefont{Georges and Kotliar}(1992)}]{GeorgesKotliar1992}
\bibinfo{author}{\bibfnamefont{A.}~\bibnamefont{Georges}} \bibnamefont{and}
  \bibinfo{author}{\bibfnamefont{G.}~\bibnamefont{Kotliar}},
  \bibinfo{journal}{Phys. Rev. B} \textbf{\bibinfo{volume}{45}},
  \bibinfo{pages}{6479} (\bibinfo{year}{1992}).

\bibitem[{\citenamefont{Eckstein and
  Werner}(2011{\natexlab{b}})}]{EcksteinWerner2011b}
\bibinfo{author}{\bibfnamefont{M.}~\bibnamefont{Eckstein}} \bibnamefont{and}
  \bibinfo{author}{\bibfnamefont{P.}~\bibnamefont{Werner}},
  \bibinfo{journal}{Phys. Rev. Lett.} \textbf{\bibinfo{volume}{107}},
  \bibinfo{pages}{186406} (\bibinfo{year}{2011}{\natexlab{b}}).

\bibitem[{Mom()}]{MomentumDistribution}
\bibinfo{note}{$k$ is defined such that $\epsilon_{\bm k}=-\cos k$ and $v_{\bm
  k}\equiv\sum_i \frac{\partial \epsilon_{\bm k}}{\partial k_i}=\sin k$
  ($-\pi\le k\le \pi$).}

\bibitem[{Hau()}]{Hauri}
\bibinfo{note}{C. Hauri, private communication; S. Watanabe, N. Minami, and R.
  Shimano, Opt. Exp. {\bf 19}, 1528 (2011).}

\bibitem[{\citenamefont{Perfetti et~al.}(2007)\citenamefont{Perfetti, Loukakos,
  Lisowski, Bovensiepen, Eisaki, and Wolf}}]{Perfetti2007}
\bibinfo{author}{\bibfnamefont{L.}~\bibnamefont{Perfetti}},
  \bibinfo{author}{\bibfnamefont{P.~A.} \bibnamefont{Loukakos}},
  \bibinfo{author}{\bibfnamefont{M.}~\bibnamefont{Lisowski}},
  \bibinfo{author}{\bibfnamefont{U.}~\bibnamefont{Bovensiepen}},
  \bibinfo{author}{\bibfnamefont{H.}~\bibnamefont{Eisaki}}, \bibnamefont{and}
  \bibinfo{author}{\bibfnamefont{M.}~\bibnamefont{Wolf}},
  \bibinfo{journal}{Phys. Rev. Lett.} \textbf{\bibinfo{volume}{99}},
  \bibinfo{pages}{197001} (\bibinfo{year}{2007}).

\bibitem[{\citenamefont{Mryasov and Freeman}(2001)}]{MryasovFreeman2001}
\bibinfo{author}{\bibfnamefont{O.~N.} \bibnamefont{Mryasov}} \bibnamefont{and}
  \bibinfo{author}{\bibfnamefont{A.~J.} \bibnamefont{Freeman}},
  \bibinfo{journal}{Phys. Rev. B} \textbf{\bibinfo{volume}{64}},
  \bibinfo{pages}{233111} (\bibinfo{year}{2001}).

\bibitem[{\citenamefont{Arita et~al.}(2004)\citenamefont{Arita, Miyake, Kotani,
  Schilfgaarde, Oka, Kuroki, Nozue, and Aoki}}]{Arita2004}
\bibinfo{author}{\bibfnamefont{R.}~\bibnamefont{Arita}},
  \bibinfo{author}{\bibfnamefont{T.}~\bibnamefont{Miyake}},
  \bibinfo{author}{\bibfnamefont{T.}~\bibnamefont{Kotani}},
  \bibinfo{author}{\bibfnamefont{M.~v.} \bibnamefont{Schilfgaarde}},
  \bibinfo{author}{\bibfnamefont{T.}~\bibnamefont{Oka}},
  \bibinfo{author}{\bibfnamefont{K.}~\bibnamefont{Kuroki}},
  \bibinfo{author}{\bibfnamefont{Y.}~\bibnamefont{Nozue}}, \bibnamefont{and}
  \bibinfo{author}{\bibfnamefont{H.}~\bibnamefont{Aoki}},
  \bibinfo{journal}{Phys. Rev. B} \textbf{\bibinfo{volume}{69}},
  \bibinfo{pages}{195106} (\bibinfo{year}{2004}).

\bibitem[{\citenamefont{Nakamura et~al.}(2009)\citenamefont{Nakamura,
  Koretsune, and Arita}}]{NakamuraKoretsuneArita2009}
\bibinfo{author}{\bibfnamefont{K.}~\bibnamefont{Nakamura}},
  \bibinfo{author}{\bibfnamefont{T.}~\bibnamefont{Koretsune}},
  \bibnamefont{and} \bibinfo{author}{\bibfnamefont{R.}~\bibnamefont{Arita}},
  \bibinfo{journal}{Phys. Rev. B} \textbf{\bibinfo{volume}{80}},
  \bibinfo{pages}{174420} (\bibinfo{year}{2009}).

\end{thebibliography}

\end{document}